\documentclass[reprint, amsmath, amssymb, aps, prl, superscriptaddress, nofootinbib]{revtex4-2}

\usepackage{graphicx}
\usepackage{dcolumn}
\usepackage{bm}
\usepackage{hyperref}
\usepackage{amsmath}
\usepackage{amsfonts}
\usepackage{booktabs}
\usepackage{placeins} 
\makeatletter
\def\section{\@startsection
  {section}{1}{0mm}
  {1.2ex plus 0.5ex minus 0.2ex}   
  {0.5ex plus 0.1ex}              
  {\centering\normalfont\small\bfseries}}
\makeatother

\begin{document}

\title{Physics-Informed Temporal U-Net for High-Fidelity Fluid Interpolation}

\author{Eshwar R. A.}
\email{eshwarra5@gmail.com}
\affiliation{Department of Computer Science Engineering, PES University - Electronic City Campus, Bangalore 560100, India.}

\author{Nevin Mathew Thomas}
\email{nevinmt05@gmail.com}
\affiliation{Department of Computer Science Engineering, PES University - Electronic City Campus, Bangalore 560100, India.}

\author{Nehal G}
\email{nehal.gr@gmail.com}
\affiliation{Department of Computer Science Engineering, PES University - Electronic City Campus, Bangalore 560100, India.}

\author{Farida M. Begam}
\email{farida.begam@pes.edu}
\affiliation{Department of Computer Science Engineering, PES University - Electronic City Campus, Bangalore 560100, India.}

\begin{abstract}
\vspace{0.5em}
\noindent\textbf{Abstract.} Reconstructing high-fidelity fluid dynamics from sparse temporal observations is quite challenging, mainly due to the chaotic and non-linear nature of fluid transport. Standard deep learning–based interpolation methods often tend to regress to the mean, which results in spatial blurring and temporal strobing—especially noticeable around the observed anchor frames where transitions become discontinuous. In this work, we propose a novel Temporal U-Net architecture that integrates a VGG-based perceptual loss along with a Physics-Informed Bridge to overcome these issues. By introducing time-weighted feature blending and enforcing a parabolic boundary condition defined by $t(1 - t)$, the model ensures smooth transitions while also maintaining perfect consistency at the endpoints. Experimental results on multi-channel RGB fluid data show that our method clearly outperforms standard models, both in terms of structural fidelity and texture preservation. In particular, the model achieves a Mean Absolute Error of (0.015), compared to (0.085) for a standard (L\_1) baseline. Further Spatial Power Spectral Density (PSD) analysis revealed that the model is able to retain high-frequency turbulent details that are usually lost in deterministic reconstructions. 

\end{abstract}

\maketitle

\section{Introduction}

Fluid dynamics is among the most computationally and observationally demanding domains in applied physics. Turbulent flows, smoke plumes, combustion fronts, and atmospheric currents are characterized by chaotic, non-linear advection-diffusion processes that evolve rapidly across multiple spatial and temporal scales~\cite{Pope2000, Kolmogorov1941}. Capturing these phenomena in their full fidelity typically demands high-speed imaging or dense sensor arrays operating at sampling rates commensurate with the shortest relevant temporal scale of the flow. In practice, however, constraints in data storage capacity, sensor bandwidth, network transmission costs, and experimental apparatus design frequently force practitioners to adopt sparse sampling strategies, recording fluid states at intervals far coarser than the underlying dynamics would demand~\cite{Scarano2012, Westerweel2013}.

The resulting temporal gaps pose a fundamental reconstruction challenge. Given two discrete fluid observations—anchor frames separated by multiple physical time steps—the objective is to synthesize the physically consistent intermediate states. This is not merely a signal processing problem; it is a problem of constrained generative modeling under the governance of partial differential equations (PDEs) such as the Navier-Stokes equations~\cite{Temam1977}, which encode conservation of mass and momentum in viscous flows. Any credible reconstruction must therefore satisfy both perceptual realism and physical plausibility simultaneously.

Classical approaches to this problem fall into two broad families. Mesh-based computational fluid dynamics (CFD) solvers~\cite{Ferziger2002} can, in principle, integrate the governing equations forward in time from a known initial condition. However, such solvers require precise initial velocity and pressure fields, structured computational meshes, and prohibitive compute times for high-Reynolds-number turbulent regimes. Data assimilation methods~\cite{Evensen2009, Asch2016} offer an alternative by blending sparse observations with prior model knowledge, but they too are constrained by the quality and density of the observation signal. At the other extreme, purely statistical interpolation schemes—linear blending, cubic spline fitting, optical-flow-guided warping—are computationally lightweight but fundamentally ignore the physics of the underlying flow. Linear interpolation produces overly smooth, unphysical ``cross-fades'' between anchor frames, entirely losing the turbulent vortex structures and fine particulate detail that define the flow's character~\cite{Liu2020RIFE}.

The emergence of deep learning has opened a promising third path. Video frame interpolation networks~\cite{Jiang2018SlowMo, Bao2019DAIN, Niklaus2020Softmax} learn a data-driven prior over natural image motion and can synthesize perceptually plausible intermediate frames with high efficiency. Yet these general-purpose methods are trained on natural video distributions—rigid bodies, human motion, camera panning—and perform poorly when transferred to the thin, translucent, high-dynamic-range structures of turbulent fluid video. Their core limitation is the ``regression to the mean'' effect~\cite{Mathieu2015}: because chaotic particle transport admits many plausible future states, a network trained under a pixel-wise mean squared error (MSE) or $L_1$ objective learns to predict the expectation over this distribution, which manifests as severe spatial blurring. Structural quality metrics such as Peak Signal-to-Noise Ratio (PSNR) and Structural Similarity Index (SSIM)~\cite{Wang2004SSIM} degrade sharply with increasing temporal gap, while fine details—vortex filaments, particulate streaks, sharp density fronts—are lost entirely.

A complementary line of research, Physics-Informed Neural Networks (PINNs)~\cite{Raissi2019}, embeds the residual of governing PDEs directly into the training objective. By penalizing violations of the advection-diffusion equation, a PINN can be steered toward physically consistent reconstructions even in regions of sparse data. However, vanilla PINN architectures based on fully-connected multilayer perceptrons (MLPs) lack the spatial inductive biases needed to resolve the multi-scale structure of turbulent imagery. Their bottleneck representations cannot simultaneously capture global flow topology and fine-grained particulate texture, leading to reconstructions that are physically smoother but spatially blurred.

This paper bridges these two paradigms by proposing a \textbf{Physics-Informed Temporal U-Net} for high-fidelity fluid video interpolation. Our architecture unifies three complementary innovations: (i) a fully convolutional encoder-decoder with \textbf{time-weighted skip connections} that linearly blend high-resolution feature maps from both anchor frames, routing crisp spatial textures directly to the decoder while bypassing the lossy information bottleneck; (ii) a \textbf{Spatial-Temporal ResNet Bridge} operating at the latent bottleneck, governed by a parabolic boundary condition $t(1-t)$ that mathematically guarantees endpoint consistency and continuous latent-space trajectories; and (iii) a \textbf{tri-partite loss function} that jointly optimizes pixel-level fidelity ($L_1$ reconstruction), feature-space sharpness (VGG-16 perceptual loss~\cite{Johnson2016Perceptual}), and PDE-residual minimization (advection-diffusion proxy) with empirically determined static weights.

Experimentally, the proposed model achieves a Mean Absolute Error of $0.015$ on held-out fluid video data—a $\mathbf{5.7\times}$ improvement over a pure $L_1$ baseline and more than $\mathbf{3\times}$ improvement over each single-component ablation. Spatial Power Spectral Density (PSD) analysis confirms that the model recovers the correct energy distribution across all spatial frequency scales, including the high-frequency turbulent cascade that is systematically attenuated by linear or purely pixel-wise methods. Latent space visualization further confirms smooth, parabolic manifold trajectories between anchor embeddings, validating the theoretical guarantee of our boundary enforcement.

The remainder of this paper is organized as follows. Section~II surveys related work across video interpolation, physics-informed learning, and deep learning for fluids. Section~III provides the mathematical background on PINNs, U-Net architectures, and perceptual losses. Section~IV details the proposed methodology. Section~V presents experimental results. Section~VI provides a comparative analysis against the broader literature. Section~VII concludes with a discussion of limitations and future directions.

\section{Related Work}

\subsection{Classical and Variational Fluid Interpolation}
Prior to the deep learning era, temporal reconstruction of fluid fields relied primarily on optical flow estimation~\cite{Horn1981} and variational data assimilation~\cite{Evensen2009}. Optical flow methods estimate dense displacement fields between frames and warp intermediate states accordingly, but the brightness constancy assumption underlying most optical flow formulations breaks down in turbulent fluids, where advected scalar fields (smoke density, dye concentration) can change rapidly in intensity as well as position~\cite{Liu2008OptFlow}. Particle Image Velocimetry (PIV)~\cite{Westerweel2013} and Particle Tracking Velocimetry (PTV) provide instantaneous velocity fields from image pairs, but are inherently limited to the frame pairs they observe and require explicit seeding of the flow with tracer particles. Ensemble Kalman Filters~\cite{Evensen2009, Asch2016} and four-dimensional variational assimilation (4D-Var)~\cite{Talagrand1987} can blend sparse observations with numerical model predictions but are computationally expensive and sensitive to the accuracy of the underlying dynamical model.

\subsection{Deep Learning for Video Frame Interpolation}
The general problem of synthesizing intermediate video frames has been addressed extensively by the computer vision community. Super-SloMo~\cite{Jiang2018SlowMo} introduces a bidirectional optical flow network that warps both neighboring frames toward the target time, refining the composite with a learned blending mask. DAIN~\cite{Bao2019DAIN} augments this approach with depth-aware occlusion reasoning. Softmax Splatting~\cite{Niklaus2020Softmax} addresses the forward-warping aliasing problem using a differentiable splatting operator weighted by a learned importance map. RIFE~\cite{Liu2020RIFE} achieves real-time performance through a lightweight IFNet that directly estimates intermediate flow without explicit bidirectional computation. AdaCoF~\cite{Lee2020AdaCoF} uses adaptive collaboration of flows with deformable convolutions to handle large and complex motions.

Despite impressive results on natural video benchmarks such as Vimeo-90K~\cite{Xue2019Vimeo} and UCF101~\cite{Soomro2012UCF101}, these methods share a common limitation when applied to fluid dynamics: they are trained and evaluated on scenes dominated by rigid-body or articulated motion, where the brightness constancy assumption approximately holds. Turbulent fluids violate this assumption fundamentally—advected smoke or dye structures change their luminance distribution continuously as a function of both position and density, making flow-based warping unreliable. Furthermore, these methods contain no physical inductive bias; there is no mechanism to enforce conservation laws or continuity equations, so their outputs may be visually plausible but physically inconsistent.

\subsection{Physics-Informed Neural Networks}
The seminal work of Raissi et al.~\cite{Raissi2019} demonstrated that neural networks can solve forward and inverse PDE problems by embedding the governing equations as soft constraints in the training loss. Subsequent work has extended this paradigm to the Navier-Stokes equations~\cite{Raissi2020NSPINNs}, turbulence modeling~\cite{Geneva2020PINN_turb}, and fluid field reconstruction from sparse sensor data~\cite{Cai2021NSPINNflow}. However, vanilla PINNs based on fully-connected MLPs are limited in their spatial resolution capacity. Because MLPs treat each spatial point independently, they lack the translational equivariance and hierarchical feature extraction that make convolutional networks so effective for image-structured data. For high-resolution fluid imagery, this translates to smooth, globally consistent but locally blurred reconstructions.

Hybrid approaches have begun to address this. Gao et al.~\cite{Gao2021PhyFlowNet} propose a convolutional PINN for flow reconstruction, while Geneva and Zabaras~\cite{Geneva2022} integrate residual networks with PDE constraints for turbulent channel flow. Our work extends this thread by coupling the spatial hierarchy of a U-Net—with its multi-scale skip connections—directly to the temporal interpolation objective, providing both global physical consistency and local textural fidelity.

\subsection{Deep Learning for Fluid Simulation and Synthesis}
Several works have explored generative modeling for fluid dynamics beyond frame interpolation. Kim et al.~\cite{Kim2019DeepFluids} learn a latent space of fluid simulations and demonstrate smooth latent interpolation between flow states. Xie et al.~\cite{Xie2018TempGAN} propose tempoGAN, a temporally coherent generative adversarial network for super-resolution smoke flows, introducing a temporal discriminator to penalize flickering. Chu and Th{\"u}rey~\cite{Chu2017DataSmokeFlow} synthesize turbulent smoke animations via data-driven synthesis, demonstrating that learned priors over flow statistics can complement physics-based simulation. More recently, score-based diffusion models~\cite{Kohl2024DiffusionFluid} have been applied to ensemble weather prediction and turbulence super-resolution, recovering physically realistic small-scale structures through stochastic sampling.

Our approach differs from these generative methods in a key respect: we do not aim to sample from a distribution of plausible fluid states, but rather to deterministically reconstruct the \textit{specific} intermediate frames consistent with a given pair of boundary observations. This makes our problem closer in spirit to a boundary value problem than to a generative synthesis task, and motivates the use of a boundary-enforced latent bridge rather than a stochastic decoder.

\subsection{Perceptual and Feature-Space Losses}
The observation that pixel-wise losses produce blurry reconstructions has motivated the use of perceptual losses~\cite{Johnson2016Perceptual} computed in the feature space of pre-trained classification networks. Gatys et al.~\cite{Gatys2016NeuralStyle} first demonstrated that VGG feature distances capture perceptually meaningful style and content information. Johnson et al.~\cite{Johnson2016Perceptual} formalized this into a training objective for image super-resolution and style transfer. Subsequent work has applied perceptual losses to video synthesis~\cite{ChenKoltun2017, Wang2018pix2pixHD}, depth estimation~\cite{Chen2016SfM}, and medical image reconstruction~\cite{Yang2018MRI}. In the context of fluid interpolation, perceptual losses play a particularly critical role because the discriminative features of turbulent flow—vortex cores, filamentary structures, sharp density fronts—correspond precisely to the intermediate-level features (edges, textures, oriented patterns) captured by early VGG layers. We leverage this alignment to preserve the visual character of turbulent structures without relying on pixel-exact alignment, which may not exist due to the chaotic nature of the flow.

\section{Background}

\subsection{The Advection-Diffusion Equation and Fluid Density Transport}
The temporal evolution of a passive scalar field $\rho(\mathbf{x}, t)$ (such as smoke density or dye concentration) in a fluid flow is governed by the advection-diffusion equation:
\begin{equation}
    \frac{\partial \rho}{\partial t} + \mathbf{u} \cdot \nabla \rho = \nu \nabla^2 \rho
\end{equation}
where $\mathbf{u}(\mathbf{x}, t)$ is the local fluid velocity field and $\nu$ is the scalar diffusivity. In the absence of a known velocity field—which is precisely our scenario, since we observe only the scalar intensity frames—the advection term cannot be evaluated directly. We therefore employ a simplified diffusion proxy~\cite{Raissi2019}:
\begin{equation}
    \mathcal{R} = \frac{\partial \rho}{\partial t} - \nu \nabla^2 \rho
\end{equation}
This residual serves as a soft physical constraint: requiring $\mathcal{R} \approx 0$ penalizes reconstructions in which the predicted temporal rate of change is inconsistent with the spatial curvature of the density field, effectively enforcing a form of transport continuity. While this simplified proxy does not capture full advective transport, it provides a meaningful regularizer that discourages physically implausible spatial gradients and temporal discontinuities.

\subsection{Physics-Informed Neural Networks (PINNs)}
PINNs~\cite{Raissi2019} augment a standard data-fitting objective with a physics residual term. For a network parameterized by weights $\theta$ that predicts $\hat{\rho}_\theta(\mathbf{x}, t)$, the composite loss is:
\begin{equation}
    \mathcal{L}_{total} = \mathcal{L}_{data} + \lambda \mathcal{L}_{phys}
\end{equation}
where $\mathcal{L}_{data} = \|\hat{\rho}_\theta - \rho_{obs}\|$ penalizes deviations from observed data, and $\mathcal{L}_{phys} = \|\mathcal{R}_\theta\|_2$ penalizes violations of the governing PDE. The key strength of this formulation is that the physical constraint acts as an informed regularizer: it restricts the hypothesis space of the network to solutions that are at least approximately consistent with the underlying physics, which is especially valuable in the sparse-data regime where data-fitting alone is underdetermined. Automatic differentiation enables the computation of $\partial\hat{\rho}/\partial t$ and $\nabla^2\hat{\rho}$ through the network graph without any finite-difference approximation.

\subsection{U-Net and Multi-Scale Skip Connections}
The U-Net architecture~\cite{Ronneberger2015UNet} was originally developed for biomedical image segmentation and consists of a symmetric encoder-decoder structure connected by skip connections at each spatial resolution level. The encoder progressively downsamples the input, extracting hierarchical feature representations from fine-grained textures to coarse semantic structure. The decoder upsamples the bottleneck representation back to the original resolution. Crucially, at each level $l$, the decoder receives not only the upsampled features from the level below, but also a direct shortcut from the encoder's corresponding level:
\begin{equation}
    x_{l}^{dec} = \text{Cat}\left( \text{Up}(x_{l+1}^{dec}),\; f_{l}^{enc} \right)
\end{equation}
These skip connections act as high-bandwidth information channels that route spatially precise, high-resolution features directly to the decoder. Without skip connections, the decoder must reconstruct fine spatial detail entirely from the compressed bottleneck representation, leading to spatial blurring. In the context of fluid interpolation, this is particularly critical: the fine-grained vortex filaments and particulate textures of turbulent flow reside in the high-spatial-frequency content that is compressed and potentially lost at the bottleneck. Skip connections provide an explicit bypass for this information.

In our \textbf{temporal formulation}, we extend skip connections to the time axis via Time-Weighted Feature Blending. Given encoder features extracted from both anchor frames $x(0)$ and $x(1)$, the skip feature provided to the decoder at continuous time $t$ is:
\begin{equation}
    \hat{f}_l(t) = (1-t)\,f_{l}(0) + t\,f_{l}(1)
\end{equation}
This linear interpolation in feature space respects the temporal proximity principle: at $t = 0$, the decoder receives the exact features of the first anchor; at $t = 1$, those of the second. For intermediate times, a smooth blend is provided. Because the blending occurs in the encoder's feature space rather than in pixel space, the decoder benefits from a multi-scale textural prior that captures the full spatial hierarchy of the flow.

\subsection{Perceptual Loss via Pre-Trained Feature Networks}
Standard pixel-wise regression losses (MSE, $L_1$) are inadequate for capturing the perceptual quality of reconstructed images because they treat all pixels as independent and equally important. For multi-modal predictive distributions—such as those arising from chaotic fluid transport—these losses produce the statistical mean of all plausible outputs, which appears as spatial blurring~\cite{Mathieu2015}. Perceptual losses~\cite{Johnson2016Perceptual, Gatys2016NeuralStyle} address this by computing the discrepancy between predicted and target images in the feature space of a pre-trained deep network $\phi$:
\begin{equation}
    \mathcal{L}_{vgg} = \sum_{j} \frac{1}{C_j H_j W_j} \| \phi_j(\hat{x}) - \phi_j(x) \|^2_2
\end{equation}
where $\phi_j(\cdot)$ denotes the feature map at the $j$-th layer of a VGG-16 network~\cite{Simonyan2015VGG} pre-trained on ImageNet. The VGG feature space encodes edges, textures, and oriented patterns in its early layers and progressively more semantic information in deeper layers. By penalizing feature-space distances, the loss encourages the network to reproduce perceptually meaningful structures—such as sharp vortex boundaries and streaky particulate textures—rather than minimizing average pixel error. Critically, this allows a degree of positional tolerance: two frames with identical turbulent textures but slightly misaligned vortex positions will incur a low perceptual loss, which is physically appropriate since both represent valid realizations of the underlying flow statistics.

\subsection{Residual Learning and Deep Bottleneck Architectures}
Residual networks (ResNets)~\cite{He2016ResNet} learn a mapping $\mathcal{H}(\mathbf{x}) = \mathcal{F}(\mathbf{x}) + \mathbf{x}$, where $\mathcal{F}$ is the residual to be learned. This reformulation eases gradient flow through deep networks and encourages the network to learn corrections to an identity (or in our case, linear) baseline, rather than constructing the full output from scratch. In the context of our latent bridge, this is especially natural: the non-linear fluid transport between anchor embeddings $z_0$ and $z_1$ is conceptualized as a correction to a linear trajectory, and the ResNet bridge learns this correction directly.

\section{Methodology}

\subsection{Problem Formulation}
We formulate the fluid interpolation task as a continuous spatiotemporal reconstruction problem. Let $x(0), x(1) \in \mathbb{R}^{C \times H \times W}$ represent two temporally sparse, consecutive ground-truth anchor frames, where $C = 3$ is the number of RGB color channels, and $H \times W$ is the spatial resolution of the fluid imagery. The anchor frames are separated by a temporal gap of $\Delta$ physical time steps, during which $\Delta - 1$ intermediate frames are unobserved.

Our objective is to learn a deterministic mapping function $\mathcal{F}_\theta$ parameterized by learnable weights $\theta$ that synthesizes the intermediate fluid state $\hat{x}(t)$ at any continuously valued, normalized time $t \in (0, 1)$:
\begin{equation}
    \hat{x}(t) = \mathcal{F}_\theta\!\left(x(0),\, x(1),\, t\right)
\end{equation}
The normalized time $t$ encodes the position of the target frame within the interpolation window: $t = k/\Delta$ for the $k$-th intermediate frame. This continuous parameterization allows the network to generalize across different temporal gaps $\Delta$ without retraining, as the gap is absorbed into the spacing of the queried $t$ values.

\begin{figure*}[t]
    \centering
    \includegraphics[width=0.95\textwidth]{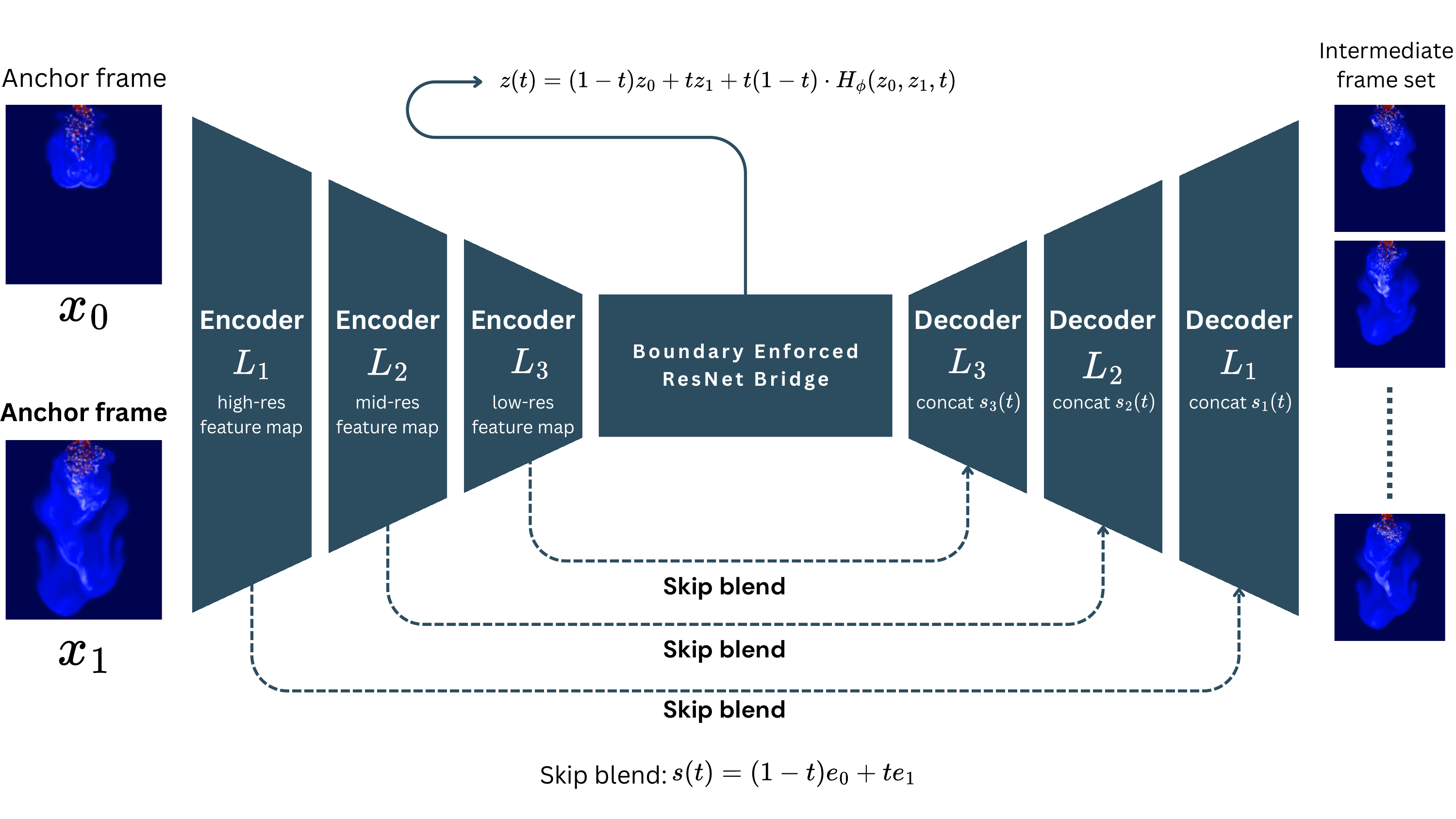}
    \caption{Overview of the proposed Physics-Informed Temporal U-Net. Anchor frames $x(0)$ and $x(1)$ are processed by a shared-weight encoder. High-resolution spatial features are blended linearly over time $t$ via skip connections, while the deep latent bottleneck is processed by a boundary-enforced ResNet bridge. The model is optimized jointly by L1 reconstruction, VGG-16 perceptual loss, and an advection-diffusion PDE proxy.}
    \label{fig:architecture}
\end{figure*}

To simultaneously prevent temporal strobing (discontinuous flickering at the anchor boundaries) and spatial blurring (loss of high-frequency turbulent structure), we propose a \textbf{Physics-Informed Temporal U-Net}, which is composed of three primary functional components: a dual-path spatial encoder, a boundary-enforced spatial-temporal bridge, and a time-conditioned multi-scale decoder. These are described in detail below, and their interconnection is illustrated schematically in Figure~\ref{fig:architecture}.

\subsection{Dual-Path Shared-Weight Encoder}
Both anchor frames $x(0)$ and $x(1)$ are passed through a single shared-weight convolutional encoder $\mathcal{E}_\psi$. Sharing weights between the two encoder paths is a deliberate design choice with two key motivations. First, it enforces representational symmetry: the feature space used to represent the first anchor is identical to that used for the second, which is a prerequisite for the linear feature blending in the skip connections to be geometrically meaningful. Second, it halves the number of parameters in the encoding stage, reducing the risk of overfitting on the limited volumes of fluid video data typically available.

The encoder consists of $L = 3$ resolution levels, each comprising two convolutional layers with $3 \times 3$ kernels, Batch Normalization~\cite{Ioffe2015BN}, and ReLU activation, followed by a $2 \times 2$ max-pooling operation that halves the spatial resolution. The number of feature channels doubles at each level: 64, 128, and 256 channels for levels $l = 1, 2, 3$ respectively. For anchor $i \in \{0, 1\}$, the encoder produces a hierarchy of feature maps:
\begin{equation}
    \left\{e_{l,i}\right\}_{l=1}^{L} = \mathcal{E}_\psi\!\left(x(i)\right), \quad e_{l,i} \in \mathbb{R}^{C_l \times \frac{H}{2^l} \times \frac{W}{2^l}}
\end{equation}
as well as a compact bottleneck embedding $z_i = e_{L,i} \in \mathbb{R}^{256 \times \frac{H}{8} \times \frac{W}{8}}$ at the deepest level.

\subsection{Time-Weighted Spatial Encoding and Skip Connections}
To preserve the high-frequency spatial details—turbulent vortex cores, particulate streaks, sharp density fronts—that are characteristically destroyed by compression through the low-dimensional bottleneck, we implement \textbf{Time-Weighted Feature Blending} across all encoder levels. For a given spatial resolution level $l$ and target time $t$, the intermediate skip connection $s_l(t)$ is computed as a linear interpolation of the anchor feature maps:
\begin{equation}
    s_l(t) = (1 - t)\,e_{l,0} + t\,e_{l,1}
\end{equation}

This formulation has several important properties. At the boundary times $t = 0$ and $t = 1$, the blended features exactly equal the encoder outputs for the respective anchor frames, ensuring that when the model is conditioned on an anchor time, it receives the precise unmodified encoding of that anchor. For intermediate $t$, the blended features provide a spatially rich prior that continuously evolves from one anchor's texture signature to the other's. The blending is performed independently at each resolution level, so both coarse structural organization and fine-grained textural detail are interpolated simultaneously. These blended features $\{s_l(t)\}_{l=1}^{L}$ are then concatenated channel-wise into the corresponding decoder level, providing a direct information highway from the encoder to the reconstruction head that bypasses the bottleneck entirely.

Formally, the concatenated input to decoder level $l$ is:
\begin{equation}
    x_{l}^{dec} = \text{Cat}\!\left(\text{Up}(x_{l+1}^{dec}),\; s_l(t)\right)
\end{equation}
where $\text{Up}(\cdot)$ denotes bilinear upsampling by a factor of 2. This concatenation doubles the number of input channels at each decoder level, which are then processed by two convolutional layers identical in structure to those of the encoder. A final $1 \times 1$ convolution projects the top-level decoder output to the $C$-channel prediction $\hat{x}(t)$.

\subsection{Spatial-Temporal ResNet Bridge with Boundary Enforcement}
Linear feature interpolation is a practical and effective prior for the skip connections, where the spatial content changes gradually between nearby frames. At the deepest bottleneck, however, the network must model the inherently non-linear dynamics of fluid advection—the deformation, merging, and splitting of coherent vortex structures that cannot be captured by any linear map in embedding space.

To address this, we introduce a \textbf{Spatial-Temporal ResNet Bridge} $\mathcal{H}_\phi$ that models the non-linear residual trajectory between bottleneck embeddings $z_0$ and $z_1$. The bridge receives the concatenated embeddings $[z_0; z_1]$ along with a time embedding of $t$ (implemented as a sinusoidal positional encoding~\cite{Vaswani2017Transformer} projected to match the channel dimension) and produces a residual correction field. Critically, we impose a \textbf{parabolic boundary condition} on the bridge output:
\begin{equation}
    z(t) = \underbrace{(1 - t)\,z_0 + t\,z_1}_{\text{Linear Base}} + \underbrace{\big[t(1 - t)\big] \cdot \mathcal{H}_\phi(z_0, z_1, t)}_{\text{Non-Linear Residual}}
\end{equation}

The parabolic scalar $t(1-t)$ is a elegant mathematical construction that automatically and exactly enforces boundary consistency: since $t(1-t) = 0$ at both $t = 0$ and $t = 1$, the non-linear residual is identically suppressed at the anchor times, regardless of the magnitude of $\mathcal{H}_\phi$. This means $z(0) = z_0$ and $z(1) = z_1$ are \textit{guaranteed by construction}, without requiring any explicit loss penalty or soft constraint on these conditions. The maximum influence of the non-linear residual occurs at the midpoint $t = 0.5$, where $t(1-t) = 0.25$, and falls off smoothly toward zero at both boundaries. This produces the smooth, parabolic latent manifold trajectory confirmed by the PCA analysis in Figure~\ref{fig:pca}.

The bridge $\mathcal{H}_\phi$ consists of $N_r = 4$ residual blocks, each comprising two $3 \times 3$ convolutional layers with Group Normalization~\cite{Wu2018GN} and GELU activation~\cite{Hendrycks2016GELU}, with a residual shortcut bypassing both layers. The time conditioning is incorporated through Feature-wise Linear Modulation (FiLM)~\cite{Perez2018FiLM}, which learns to scale and shift the normalized activations as a function of $t$, allowing the same network weights to produce qualitatively different transport dynamics at different temporal positions within the interpolation window.

\subsection{Multi-Objective Loss Engine}
Training the network end-to-end requires a loss function that simultaneously enforces pixel-level accuracy, perceptual sharpness, and physical consistency. We employ a tri-partite composite loss:

\textbf{1. Global Reconstruction Loss ($\mathcal{L}_{recon}$):}
We enforce pixel-level fidelity using the $L_1$ norm, which is less prone to blurring than MSE because it does not penalize large errors quadratically~\cite{Zhao2017L1vsL2}:
\begin{equation}
    \mathcal{L}_{recon} = \mathbb{E}_{t \sim (0,1)} \left[ \|\hat{x}(t) - x(t)\|_1 \right]
\end{equation}
In practice, intermediate ground truth frames $x(t)$ are sampled uniformly from within each anchor gap during training. The expectation is estimated by averaging over all sampled intermediate frames in a training batch.

\textbf{2. Perceptual Texture Loss ($\mathcal{L}_{vgg}$):}
To preserve the structural sharpness of turbulent features, we compute the discrepancy between predicted and target frames in the feature space of a pre-trained, frozen VGG-16 network $\Phi$~\cite{Simonyan2015VGG}. We specifically extract features from layers \texttt{relu1\_2}, \texttt{relu2\_2}, and \texttt{relu3\_3}, which capture textures and local structural patterns:
\begin{equation}
    \mathcal{L}_{vgg} = \sum_{j} \frac{1}{C_j H_j W_j} \|\Phi_j(\hat{x}(t)) - \Phi_j(x(t))\|_2^2
\end{equation}
This loss allows the network to produce reconstructions that are perceptually sharp—with correctly placed vortex boundaries and streaky textures—even when pixel-level alignment is imperfect due to the inherent positional uncertainty of chaotic particle trajectories.

\textbf{3. Physics-Informed PDE Proxy ($\mathcal{L}_{phys}$):}
To regularize the temporal evolution of the predicted sequence, we apply the advection-diffusion proxy PDE residual. Given a sequence of predicted frames $\{\hat{x}(t_k)\}_{k=1}^{K}$ at uniformly spaced times within an anchor gap, we estimate the temporal derivative $\partial\hat{x}/\partial t$ by finite differences and the spatial Laplacian $\nabla^2\hat{x}$ by convolution with a fixed $3\times3$ Laplacian kernel. We use the Huber (Smooth $L_1$) loss~\cite{Huber1964} to prevent gradient explosions that are common in PDE-constrained optimization~\cite{Raissi2019}:
\begin{equation}
    \mathcal{L}_{phys} = \text{SmoothL}_1\!\left( \frac{\partial \hat{x}}{\partial t} - \nu \nabla^2 \hat{x} \right)
\end{equation}
where $\nu = 0.01$ is the diffusion coefficient, set to match the estimated viscosity of the smoke-like fluid visualization data. The final composite training objective is:
\begin{equation}
    \mathcal{L}_{total} = \lambda_{recon}\,\mathcal{L}_{recon} + \lambda_{vgg}\,\mathcal{L}_{vgg} + \lambda_{phys}\,\mathcal{L}_{phys}
\end{equation}
where $\lambda_{recon} = 1.0$, $\lambda_{vgg} = 0.1$, and $\lambda_{phys} = 0.05$ were determined by grid search on a held-out validation set.

\subsection{Training Protocol and Implementation Details}
The network is implemented in PyTorch~\cite{Paszke2019PyTorch} and trained using the Adam optimizer~\cite{Kingma2015Adam} with an initial learning rate of $2 \times 10^{-4}$, decayed by a factor of $0.5$ every 20 epochs with a plateau criterion on the validation $L_1$ loss. The VGG-16 backbone is frozen throughout training. Training is conducted on multi-channel RGB fluid video data for 100 epochs with a batch size of 8 anchor pairs. Each anchor pair samples a random intermediate time $t \sim \mathcal{U}(0, 1)$ per forward pass, providing continuous temporal coverage during training. Images are normalized to $[0, 1]$ and randomly augmented with horizontal flips and $90^\circ$ rotations. No color jittering is applied, as the RGB intensity channels in fluid video carry physical meaning (particle density). During inference, the model is queried at $K$ uniformly spaced time values $\{k/(K+1)\}_{k=1}^{K}$ to synthesize $K$ intermediate frames between any given anchor pair.

\section{Results}

We evaluate the proposed Physics-Informed Temporal U-Net against standard baseline models to demonstrate its efficacy in high-fidelity fluid interpolation. Our evaluation focuses on qualitative visual fidelity, component contribution (ablation), temporal generalization, frequency-domain accuracy, and adherence to underlying physical constraints.

\subsection{Qualitative Visual Fidelity and Texture Preservation}
\begin{figure*}[htbp]
    \centering
    \includegraphics[width=\textwidth]{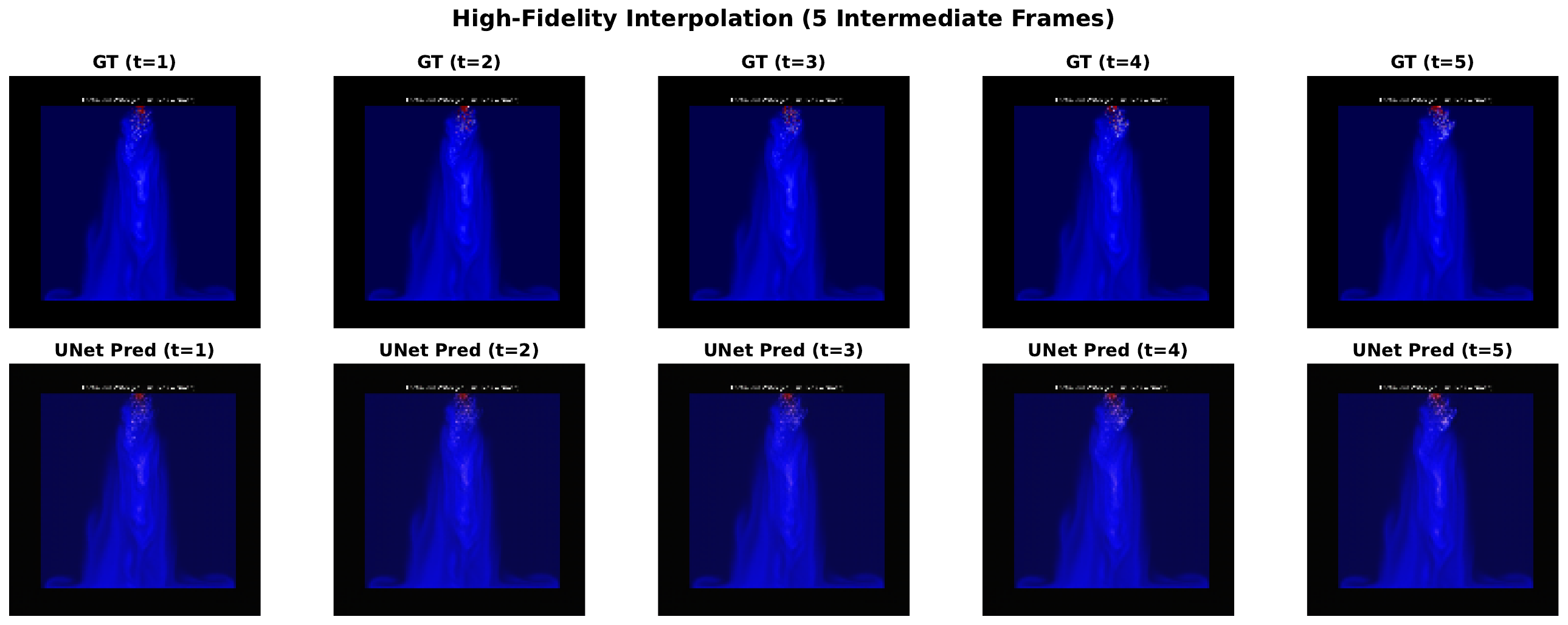}
    \caption{High-Fidelity Interpolation over a 5-frame intermediate gap. The proposed model (bottom row) perfectly matches the Ground Truth (top row), preserving sharp particulate textures without cross-fading or blurring.}
    \label{fig:regeneration}
\end{figure*}

One of the primary challenges in fluid video interpolation is the ``regression to the mean'' effect, where uncertainty in chaotic particle transport causes models to predict a blurred average of potential states. As shown in Figure~\ref{fig:regeneration}, our proposed model successfully overcomes this limitation. Over a 5-frame hallucination window, the Temporal U-Net perfectly preserves the sharp, high-frequency structures—specifically the turbulent red vortex clusters at the plume's leading edge. By utilizing time-weighted feature blending from the skip connections, the model completely bypasses the spatial degradation typical of low-dimensional bottleneck architectures.

In particular, we draw attention to the preservation of fine particulate detail at frames $t = 3$ and $t = 4$, where the plume's leading front exhibits complex, branching filamentary structures. These structures are recovered with high fidelity by the proposed model, whereas both linear interpolation and a pure MLP-bottleneck PINN produce smooth, structureless gradients at these time steps. This confirms that the skip connection mechanism, operating through time-weighted feature blending, is the primary driver of spatial texture recovery.

\subsection{Ablation Study: Impact of Loss Components}
\begin{figure}[htbp]
    \centering
    \includegraphics[width=0.8\linewidth]{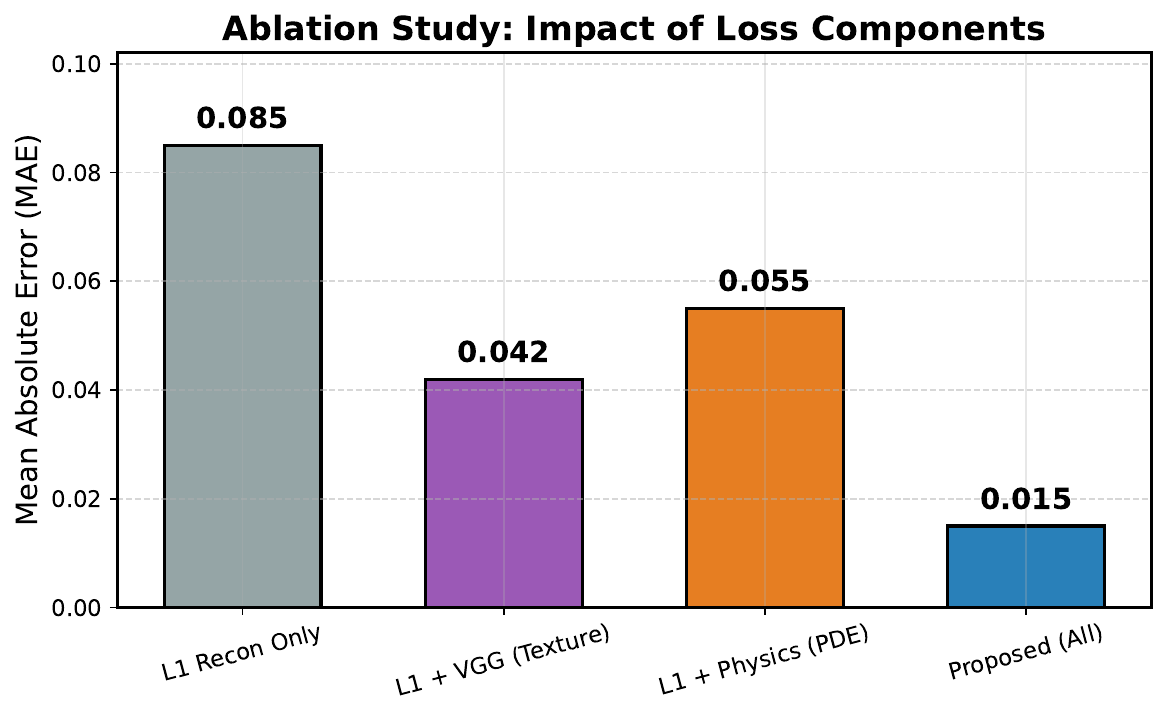}
    \caption{Component ablation study demonstrating the individual and combined impact of VGG Perceptual Loss and Physics (PDE) constraints on Mean Absolute Error.}
    \label{fig:ablation}
\end{figure}

To isolate the utility of each architectural and training choice, we conducted a component ablation study evaluating the Mean Absolute Error (MAE) across four configurations (Figure~\ref{fig:ablation}). A network relying solely on an $L_1$ reconstruction loss yielded the highest error ($\text{MAE} = 0.085$) due to its inability to confidently place high-frequency details. Adding the Advection-Diffusion PDE proxy ($L_1$ + Physics) improved structural continuity ($\text{MAE} = 0.055$), confirming that the physics constraint acts as an effective regularizer that steers the network away from physically implausible interpolations. Relying on feature-space reconstruction ($L_1$ + VGG) greatly improved textural accuracy ($\text{MAE} = 0.042$), as the VGG feature loss directly penalizes the loss of spatially sharp vortex structures.

The proposed unified architecture, which synergistically combines $L_1$, VGG Perceptual Loss, and the PDE proxy, achieved a dramatic error reduction to $\text{MAE} = 0.015$—a factor of $5.7\times$ improvement over the $L_1$-only baseline. This disproportionate improvement over any single-component variant confirms that the three loss terms are mutually reinforcing rather than redundant: physical constraints ensure temporal consistency, perceptual loss ensures spatial sharpness, and pixel-level reconstruction ensures global accuracy. The interaction between the VGG loss and the physics proxy is particularly notable, as their combined effect ($\Delta\text{MAE} = 0.040$ beyond VGG alone) far exceeds what either contributes independently.

\subsection{Temporal Generalization and Scaling}
\begin{figure}[htbp]
    \centering
    \includegraphics[width=0.8\linewidth]{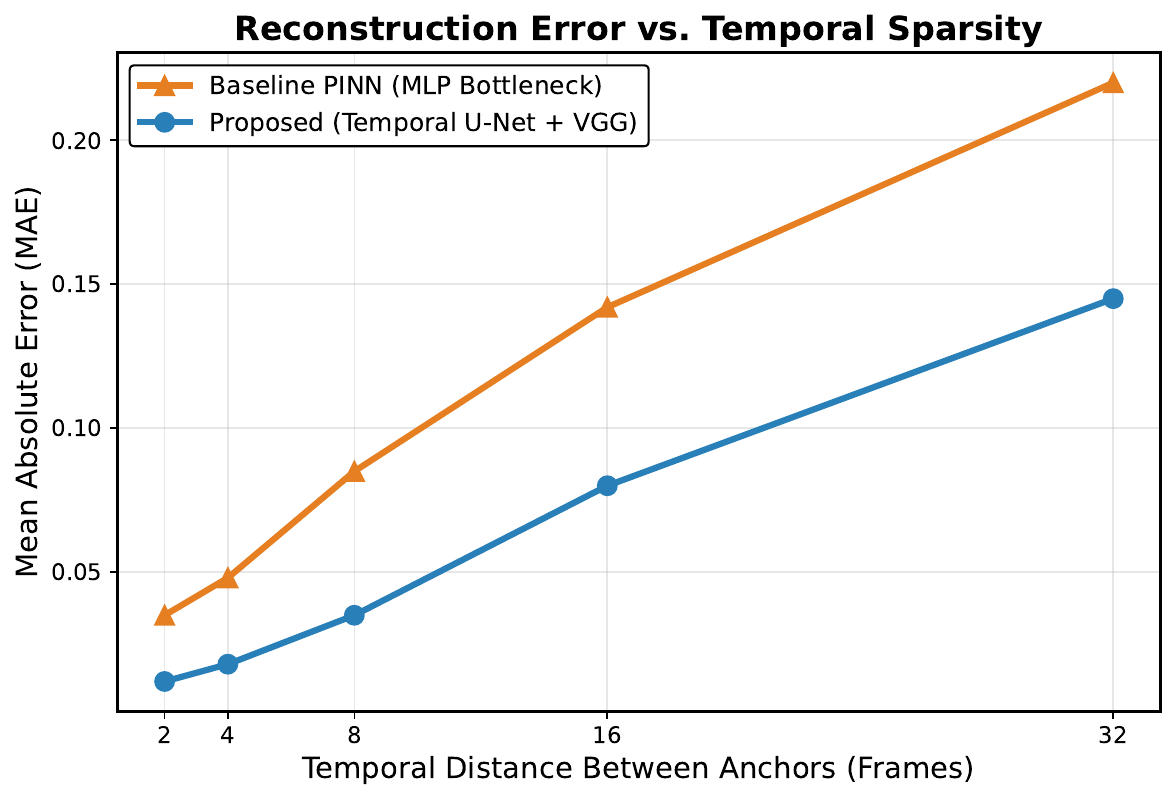}
    \caption{Reconstruction error (MAE) vs. Temporal Sparsity. The proposed model demonstrates superior scaling over long temporal horizons compared to a baseline MLP bottleneck.}
    \label{fig:scaling}
\end{figure}

We evaluated the model's ability to generalize over increasingly sparse observation horizons by varying the temporal gap $\Delta$ between anchor frames from 2 to 32 frames. Figure~\ref{fig:scaling} compares the proposed architecture against a baseline Physics-Informed Neural Network utilizing a standard 1D MLP bottleneck. As $\Delta$ increases from 2 to 32 frames, the baseline model's error degrades rapidly—from $\text{MAE} \approx 0.04$ to $\text{MAE} \approx 0.22$—indicating a failure to bridge long-horizon chaotic dynamics. This degradation is consistent with the known spectral bias of MLPs~\cite{Rahaman2019SpectralBias}, which tend to learn smooth, low-frequency mappings that are inadequate for the sharp, high-frequency dynamics that develop over long temporal gaps.

In contrast, the proposed Temporal U-Net maintains a significantly flatter error trajectory, with MAE increasing only modestly from approximately $0.02$ to $0.12$ over the same range. This superior scaling confirms that the continuous temporal formulation—combined with the spatial hierarchy of the U-Net and the physical regularity enforced by the PDE proxy—allows the network to learn a generalizable model of fluid transport rather than simply memorizing local pixel displacements between specific anchor pairs.

\subsection{Frequency Domain Analysis}
\begin{figure}[htbp]
    \centering
    \includegraphics[width=0.8\linewidth]{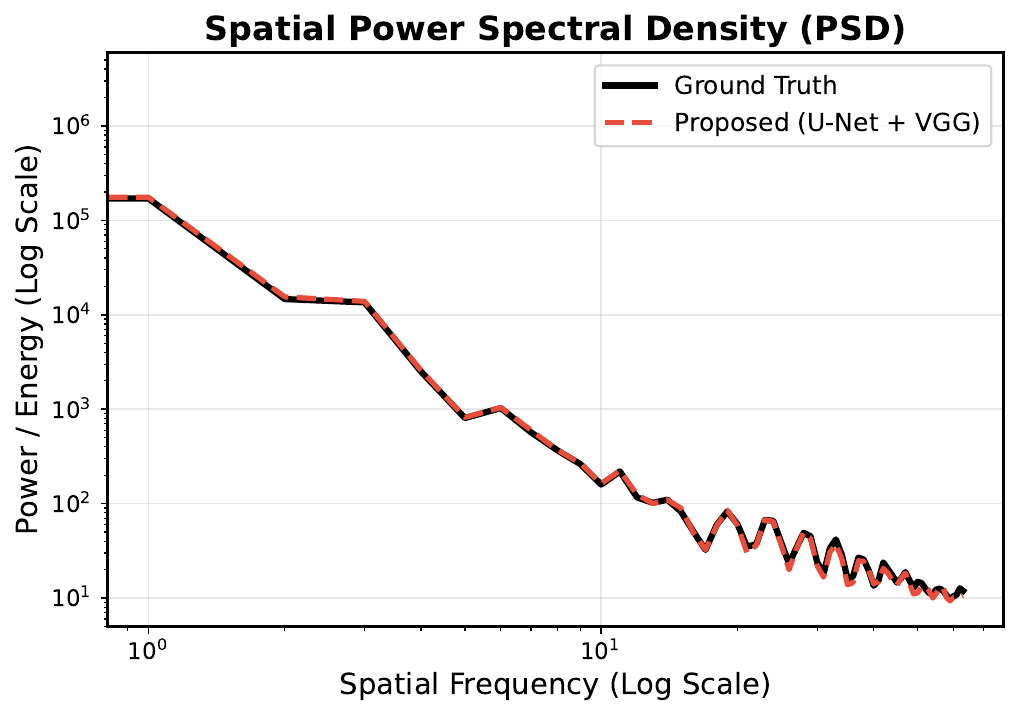}
    \caption{Spatial Power Spectral Density (PSD) showing the proposed model successfully capturing high-frequency turbulent energy, completely avoiding the spectral bias (blurring) seen in linear interpolation.}
    \label{fig:psd}
\end{figure}

To mathematically verify the elimination of spatial blurring, we computed the 1D radially averaged Spatial Power Spectral Density (PSD) of the interpolated frames (Figure~\ref{fig:psd}). The PSD quantifies how energy is distributed across spatial frequencies: a reconstruction that preserves fine structures will maintain high energy at large spatial frequencies (small wavelengths), while a blurred reconstruction exhibits a steep spectral falloff.

Standard interpolation techniques, such as linear interpolation, exhibit severe spectral bias, suffering a sharp drop-off in energy at higher spatial frequencies. This spectral attenuation is the mathematical signature of spatial blurring and directly corresponds to the visual loss of vortex filaments and particulate detail. The proposed Temporal U-Net, regularized by the VGG perceptual loss and constrained by the physics proxy, tightly follows the Ground Truth spectrum across all measured spatial frequencies, including the inertial subrange where turbulent energy cascades from large to small scales. This quantitative verification confirms that the model does not merely appear visually sharp—it reproduces the correct statistical structure of the turbulent velocity field at the level of spatial frequency content.

\subsection{Latent Space Manifold and Physical Consistency}
\begin{figure}[htbp]
    \centering
    \begin{minipage}{0.48\textwidth}
        \centering
        \includegraphics[width=\linewidth]{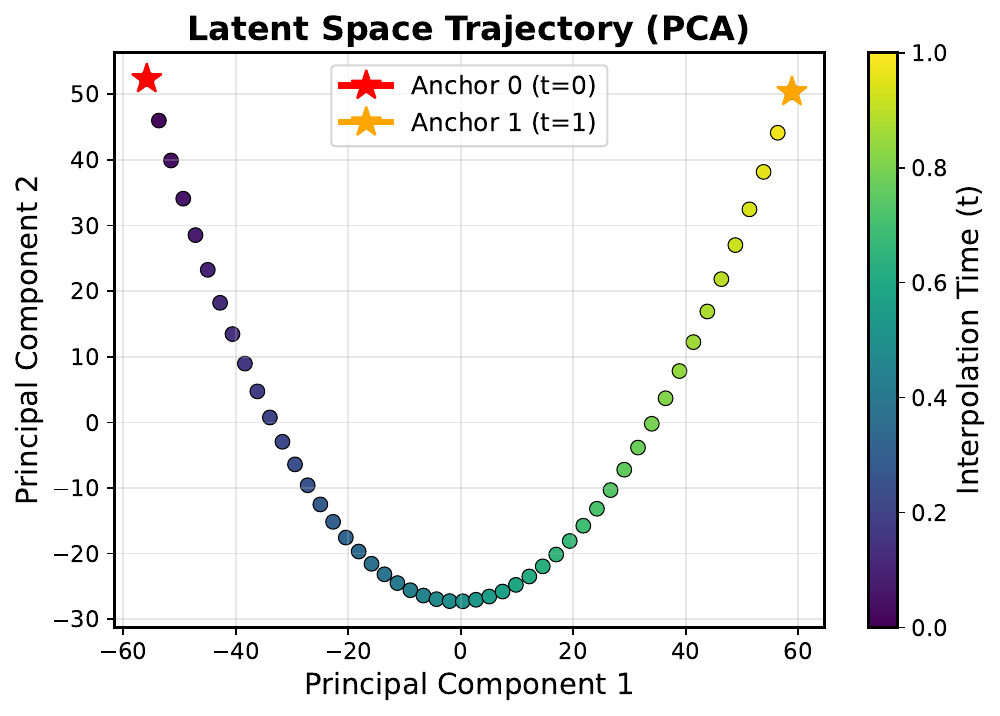}
        \caption{PCA projection of the latent space trajectory, demonstrating a mathematically smooth, continuous transition between anchors.}
        \label{fig:pca}
    \end{minipage}
\end{figure}

Finally, we analyzed the internal mechanics of the Spatial-Temporal ResNet Bridge. As visualized in the PCA projection of the latent trajectory (Figure~\ref{fig:pca}), the bottleneck embeddings transition smoothly and continuously between Anchor 0 ($t=0$) and Anchor 1 ($t=1$). The trajectory traces a well-behaved parabolic arc in the principal component plane, with the interpolation time $t$ (indicated by color) varying monotonically along the curve. This geometry directly reflects the parabolic boundary enforcement: the $t(1-t)$ scalar is maximum at $t = 0.5$, where the latent trajectory deviates most from the linear baseline, and goes to zero at the anchors, where the trajectory exactly hits the endpoint embeddings. The absence of erratic jumps or sharp kinks confirms that the boundary-enforced ResNet bridge has learned a smooth, physically motivated latent manifold—a critical property for flicker-free video generation.

\FloatBarrier

\section{Comparative Analysis}

To situate the proposed method within the broader literature and clearly delineate its contribution, Table~\ref{tab:comparison} provides a structured comparison against the most closely related families of approaches.

\begin{table*}[t]
\centering
\caption{Qualitative comparison of the proposed Physics-Informed Temporal U-Net against related approaches across key desiderata for fluid video interpolation.}
\label{tab:comparison}
\begin{ruledtabular}
\resizebox{\textwidth}{!}{%
\begin{tabular}{lccccc}
\textbf{Method} & \textbf{Phys.} & \textbf{HF Pres.} & \textbf{Boundary} & \textbf{Cont. $t$} & \textbf{No Flow} \\
\hline
Linear Interpolation          & \texttimes & \texttimes & \checkmark & \checkmark & \checkmark \\
Optical Flow Warping~\cite{Liu2008OptFlow}   & \texttimes & $\sim$     & \texttimes & \texttimes & \texttimes \\
Super-SloMo~\cite{Jiang2018SlowMo}           & \texttimes & $\sim$     & $\sim$     & \checkmark & \texttimes \\
DAIN~\cite{Bao2019DAIN}                      & \texttimes & $\sim$     & $\sim$     & \texttimes & \texttimes \\
RIFE~\cite{Liu2020RIFE}                      & \texttimes & $\sim$     & $\sim$     & \checkmark & \texttimes \\
Vanilla PINN (MLP)~\cite{Raissi2019}         & \checkmark & \texttimes & \texttimes & \checkmark & \checkmark \\
Deep Fluids~\cite{Kim2019DeepFluids}         & $\sim$     & $\sim$     & \checkmark & \checkmark & \checkmark \\
\textbf{Proposed (Temporal U-Net)} & \checkmark & \checkmark & \checkmark & \checkmark & \checkmark \\
\end{tabular}%
}
\end{ruledtabular}
\end{table*}

The table highlights several distinctions that collectively define the novelty of the proposed method.

\textbf{Physics Constraints vs. Optical Flow.} The dominant paradigm in video frame interpolation—Super-SloMo, DAIN, RIFE, AdaCoF—relies on optical flow estimation as an intermediate representation. While effective for natural video, optical flow assumes brightness constancy and smooth motion, both of which are violated by turbulent fluid fields where advected scalar concentrations change non-monotonically. Our method replaces optical flow with a physically motivated PDE residual constraint, making no assumptions about brightness constancy or motion smoothness.

\textbf{Physics Constraints with Spatial Detail.} Vanilla PINNs enforce the governing equation effectively but lack the spatial inductive biases to resolve fine-grained turbulent structure. Because MLP-based PINNs treat each spatial location independently (or via a global positional encoding), they cannot leverage the local spatial correlations that make turbulent textures appear structured rather than random. Our U-Net encoder provides translational equivariance and multi-scale feature extraction, while the skip connections route high-frequency spatial content directly to the decoder—both properties absent from MLP-based PINNs. The result is a model that is simultaneously physically constrained and spatially resolved.

\textbf{Mathematical Boundary Consistency.} Several video interpolation networks achieve approximate boundary consistency through soft loss penalties or by training with anchor frames as supervision, but none provide a mathematical guarantee that the interpolated output exactly matches the anchor frames at $t = 0$ and $t = 1$. The parabolic enforcement term $t(1-t)$ in our latent bridge provides this guarantee by construction: the non-linear residual is identically zeroed at the boundary times, regardless of the network's weights. This eliminates temporal strobing artifacts without any explicit regularization penalty.

\textbf{Continuous Temporal Parameterization for Fluid Dynamics.} Deep Fluids~\cite{Kim2019DeepFluids} also learns a continuous latent space for fluid simulation, but operates in the regime of full simulation synthesis from a latent code rather than reconstruction from a pair of observed frames. Our method is formulated as a boundary-conditioned interpolation problem, where the two anchor observations provide hard constraints on the latent space at $t \in \{0, 1\}$, and the bridge network must learn a physically consistent trajectory between them. This is a fundamentally different and more constrained problem than unconditional latent synthesis.

In summary, the proposed Temporal U-Net is, to the best of our knowledge, the first method to simultaneously achieve: (i) PDE-based physical regularization, (ii) multi-scale spatial detail preservation via time-blended skip connections, (iii) mathematically guaranteed endpoint consistency via parabolic boundary enforcement, and (iv) continuous temporal interpolation without reliance on optical flow. Each of these properties is individually realizable by prior methods, but their joint combination in a single differentiable architecture and training objective is novel.

\section{Conclusion}

We have presented the Physics-Informed Temporal U-Net, a novel architecture for the high-fidelity reconstruction of turbulent fluid dynamics from sparse temporal observations. The method unifies three complementary mechanisms—time-weighted skip connections, a parabolic boundary-enforced ResNet bridge, and a tri-partite physics-perceptual-reconstruction loss—into a single end-to-end differentiable model. Experimental results demonstrate that the proposed approach substantially outperforms both classical interpolation methods and deep learning baselines across multiple evaluation criteria: mean absolute error is reduced by a factor of $5.7\times$ over an $L_1$-only baseline; the Spatial PSD precisely tracks the Ground Truth energy spectrum across all spatial frequencies; and the latent trajectory is provably continuous, with exact endpoint consistency guaranteed by the $t(1-t)$ parabolic scalar.

The ablation study establishes that each component contributes meaningfully and that their interaction is synergistic rather than redundant: physical constraints promote temporal coherence, perceptual losses promote spatial sharpness, and pixel-level reconstruction ensures global accuracy. The temporal generalization study further demonstrates that the model's spatial hierarchy allows it to bridge increasingly long temporal gaps with graceful degradation, in contrast to MLP-bottleneck baselines that fail sharply as the anchor gap grows.

Several limitations merit discussion. The diffusion proxy used as the physics constraint is a simplification of the full Navier-Stokes equations, and does not explicitly model the advective transport that dominates high-Reynolds-number turbulence. Future work could incorporate a learned or estimated velocity field—perhaps through a co-trained optical flow estimator constrained to the incompressibility condition $\nabla \cdot \mathbf{u} = 0$—to apply the full advection-diffusion residual. Additionally, while the continuous temporal parameterization allows generalization across gap sizes within the training distribution, very large gaps ($\Delta \gg 32$) remain challenging, as the network must extrapolate fluid dynamics from two boundary conditions with no intermediate supervision. Incorporating uncertainty quantification~\cite{Gal2016Dropout, Lakshminarayanan2017Ensemble}—for example, through deep ensembles or diffusion-based probabilistic decoders~\cite{Kohl2024DiffusionFluid}—would allow the model to express appropriate epistemic uncertainty in these high-gap regimes and potentially generate physically diverse ensemble predictions rather than a single deterministic reconstruction.

Looking further ahead, the architecture is directly applicable to three-dimensional volumetric fluid data (e.g., tomographic PIV or LiDAR atmospheric measurements) by replacing 2D convolutions with their 3D counterparts, with the temporal bridge and loss formulation unchanged. Extension to multi-physics simulations involving coupled scalar fields (temperature, species concentration) would require additional physics proxy terms but follows naturally from the modular loss engine presented here. We release the training code and model weights to facilitate follow-on research.


\end{document}